# The Near-Infrared Sky Surveyor[1]

Daniel Stern (daniel.k.stern@jpl.nasa.gov)
Jet Propulsion Laboratory / California Institute of Technology

**Co-Investigators:**
James G. Bartlett (JPL/Caltech)
Mark Brodwin (Harvard/CfA)
Asantha Cooray (UC-Irvine)
Roc Cutri (IPAC/Caltech)
Arjun Dey (NOAO)
Peter Eisenhardt (JPL/Caltech)
Anthony Gonzalez (Univ. Florida)
Jason Kalirai (STScI)
Amy Mainzer (JPL/Caltech)
Leonidas Moustakas (JPL/Caltech)
Jason Rhodes (JPL/Caltech)
S. Adam Stanford (UC-Davis, IGPP)
Edward L. Wright (UCLA)

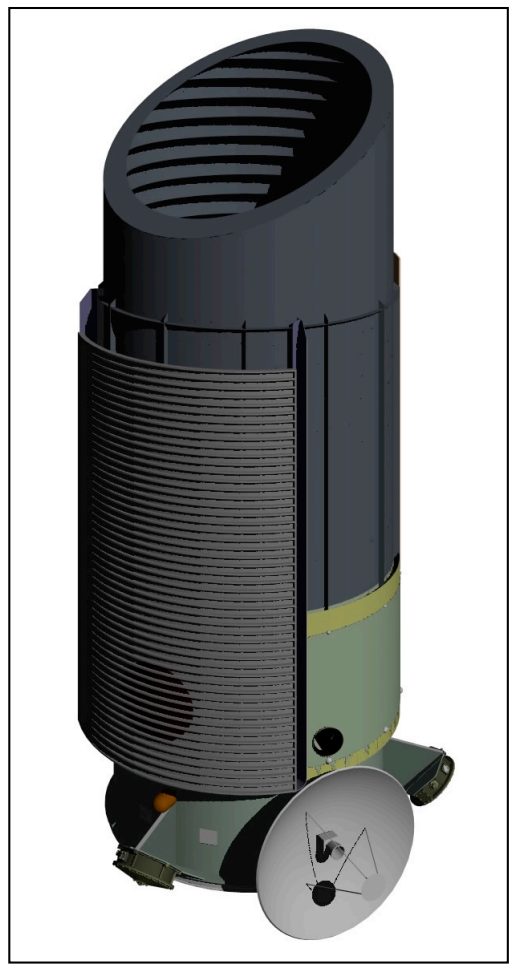

Submitted on April 1, 2009
*[budget section excised]*

A Response to the Astro2010 Request for Information
by the Electromagnetic Observations from Space (EOS) Panel

[1] *NIRSS* is one of three concepts that contributed to the *Wide-Field Infrared Survey Telescope (WFIRST)* mission advocated by the Decadal Survey.

# 1  Executive Summary

Operating beyond the reaches of the Earth's atmosphere, free of its limiting absorption and thermal background, the *Near-Infrared Sky Surveyor* (*NIRSS*) will deeply map the entire sky at near-infrared wavelengths, thereby enabling new and fundamental discoveries ranging from the identification of extrasolar planets to probing the reionization epoch by identifying thousands of quasars at $z>10$. *NIRSS* will directly address the NASA scientific objective of studying cosmic origins by using a 1.5-meter telescope to reach full-sky 0.2 µJy (25.6 mag AB) sensitivities in four passbands from 1 to 4 µm in a 4-yr mission. At the three shorter passbands ($1-2.5$ µm), the proposed depth is comparable to the deepest pencil-beam surveys done to date and is ~3000 times more sensitive than the only previous all-sky near-infrared survey, 2MASS (Fig. 1). At the longest passband (3.5 µm), which is not feasible from the ground, *NIRSS* will be 500 times more sensitive than *WISE*. *NIRSS* fills a pivotal gap in our knowledge of the celestial sphere, is a natural complement to *WISE,* and is well matched to the next generation of deep (~0.1 µJy), wide-area (>2π ster), ground-based optical surveys (LSST and Pan-Starrs; Fig. 2). With the high thermal backgrounds of ground-based infrared observations, a near-infrared full sky survey at sub-µJy sensitivity is only feasible from space.

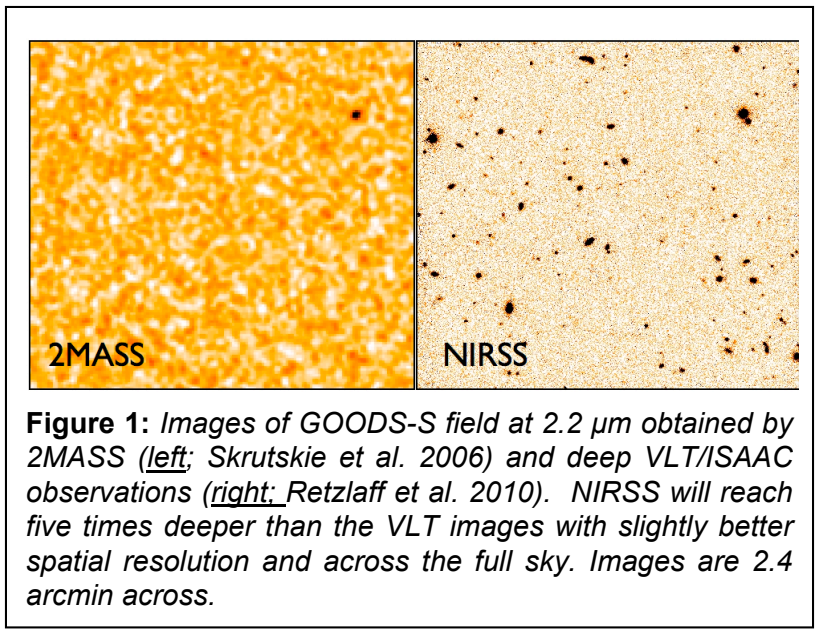

**Figure 1:** *Images of GOODS-S field at 2.2 µm obtained by 2MASS (left; Skrutskie et al. 2006) and deep VLT/ISAAC observations (right; Retzlaff et al. 2010). NIRSS will reach five times deeper than the VLT images with slightly better spatial resolution and across the full sky. Images are 2.4 arcmin across.*

Shallow, ground-based, near-infrared surveys have revolutionized our understanding of sub-stellar objects, identified populations of heavily obscured active galaxies, and probed galaxy and structure formation at intermediate distances. While ultra-deep infrared images have only recently been obtained over small areas, these data are already changing our understanding of galaxy formation. *NIRSS* will build on these achievements by providing ultra-deep near-infrared images across the entire sky. Besides its substantial legacy, *NIRSS* will:

- directly measure the diffuse extragalactic background produced by the ensemble of primordial galaxies, thereby probing the first generation of galaxy formation;
- fundamentally alter the landscape of early universe studies by identifying and studying large samples of quasars and galaxies in the first billion years after the Big Bang;
- measure the spectral energy distributions of billions of galaxies at moderate redshifts, thereby significantly enhancing weak lensing studies of the large-scale distribution of luminous and dark matter across cosmic time and directly testing theories of structure and galaxy formation;
- find the coolest Galactic sources, thereby probing both the star-planet boundary and the origins and age of the Milky Way Galaxy; and
- conduct a census of exoplanets in orbits beyond 1 AU via synoptic microlensing observations of the Galactic bulge.

## 2  Key Science Goals

*NIRSS* occupies a unique position in area-depth space relative to previous near-infrared surveys, and a unique wavelength-depth space relative to previous all-sky surveys (Fig. 2). The proposed 0.2 µJy depth (25.6 mag AB) out to 4 µm is 10 times deeper than the Vista VIDEO and UKIDSS DXS surveys, two projects which only work out to ~2.5 µm, will take several years to complete, and will only cover <0.01% of the sky. Since the cost of a large infrared focal plane is substantial, launching it into the benign thermal environment of space is much more cost effective than building multiple arrays on the ground — e.g., based on the above surveys, reaching *NIRSS* depths across the full sky in the three shorter passbands would require several hundred 4-m class telescopes, each with their own wide-area focal planes. Deep, wide-area surveys at the longest passband of *NIRSS* are simply not possible from the ground.

*NIRSS* will provide a legacy archive for long-term reference, useful for identifying targets for both the *JWST* and the next generation of 30-meter class ground-based telescopes, and will be invaluable for scientific investigations ranging from probing the diffuse background from primordial galaxies to identifying the coldest Galactic brown dwarfs. In its 4 yr mission, *NIRSS* will also be sensitive to the proper motions of faint, red Galactic objects at an unprecedented level, a powerful tool for studying the Solar neighborhood. Naturally, a survey as fundamental as *NIRSS* has a wide range of astrophysical uses; in the following subsections we highlight just four of the key investigations enabled by *NIRSS*. Additional information can also be found in scientific white papers submitted to the Astro2010 Decadal Survey by D.Bennett (microlensing exoplanet survey), A.Burgasser (brown dwarfs), A.Cooray (diffuse background), J.Kalirai (white dwarfs), J.Prochaska (high-redshift quasars) and S.A.Stanford (high-redshift galaxies).

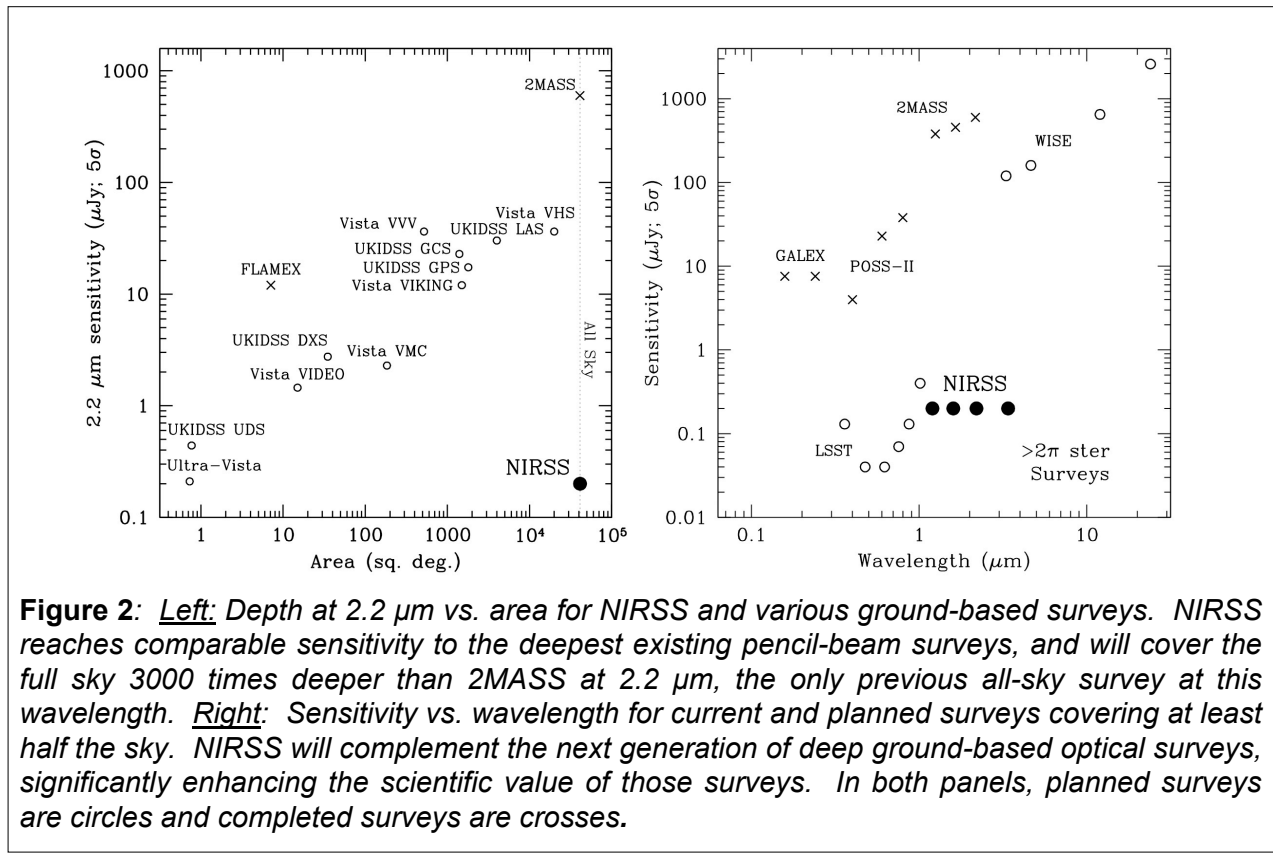

**Figure 2**: <u>Left</u>: *Depth at 2.2 µm vs. area for NIRSS and various ground-based surveys. NIRSS reaches comparable sensitivity to the deepest existing pencil-beam surveys, and will cover the full sky 3000 times deeper than 2MASS at 2.2 µm, the only previous all-sky survey at this wavelength.* <u>Right</u>: *Sensitivity vs. wavelength for current and planned surveys covering at least half the sky. NIRSS will complement the next generation of deep ground-based optical surveys, significantly enhancing the scientific value of those surveys. In both panels, planned surveys are circles and completed surveys are crosses.*

## 2.1 Diffuse Infrared Background from Population III Stars

The first galaxies cannot be individually detected with current instruments, and even *JWST* will only detect the brightest protogalaxies selected in small fields. However, the ensemble of primordial galaxies produces a diffuse extragalactic background, most prominently at near-infrared wavelengths, which will be detected by *NIRSS*. This background, which directly probes the earliest phases of star formation in the universe, has a distinct power spectrum which peaks at angular scales of ~7.5 arcmin (Cooray *et al*. 2004). Indeed, the absolute intensity of the infrared background, as measured by DIRBE and *IRTS*, exceeds integrated galaxy counts in deep fields. This implies a missing component that may be due to the first generation of stars formed after the Big Bang, *e.g.*, "*Population III stars*" (*e.g,* Santos *et al*. 2002). Analyzing data from narrow (~ 100 sq. arcmin), deep *Spitzer* fields, Kashlinsky *et al*. (2005, 2007) have recently reported detection of background fluctuations from first-light galaxies at 3.6 to 4.5 μm. These exciting, yet controversial, results would have profound cosmological implications, and have generated significant attention both in the press and in the astrophysical community. However, there is concern that a significant fraction of the reported background is due to faint, foreground sources, just below the detection limit (Cooray *et al*. 2007, Sullivan *et al*. 2007). *NIRSS* will definitively and unambiguously measure the infrared background from Population III stars.

A fundamental limitation of previous work studying primordial galaxy background fluctuations is that their limited fields, ~10 arcmin across, only probe small-scale fluctuations. The wide-area coverage of *NIRSS* is significantly better for studying the diffuse background since it probes the first-galaxy fluctuation spectrum on all scales. Importantly, the color across the *NIRSS* bands strongly discriminates first-light fluctuations from foregrounds (*e.g.*, zodiacal emission) and identifies at which cosmic epoch the first stars formed in the universe. Zodiacal emission will also have a time-dependent component and is believed to be smooth on ~30 arcmin scales. Current theoretical models put first light at redshift $z \sim 10$ to 20 (*e.g.*, Furlanetto & Loeb 2005), predicting a sharp spectral signature in the diffuse background at 1.3 to 2.6 μm. *NIRSS* will directly measure the wavelength where this background changes intensity, thus identifying the first epoch of star formation. Studying the infrared background requires the thermal stability of space, and thus is not feasible from the ground. Furthermore, other current and planned space-based infrared missions either have too small of a field of view (*e.g.*, *JWST*), work too far into the infrared (*e.g.*, *Spitzer* and *WISE*), or do not work sufficiently far into the near-infrared (*e.g.*, JDEM) to fully study the diffuse background from primordial stars.

## 2.2 The Most Distant Quasars

Quasars are among the most luminous objects in the universe, observable to the earliest cosmic epochs. They provide fundamental information on the earliest phases of structure formation in the universe and are unique probes of the intergalactic medium. The discovery of a Gunn-Peterson (1965) trough in the spectra of several quasars at redshift $z > 6$ suggests that the universe completed reionization near that redshift (*e.g.*, Fan *et al*. 2001), though poor sample statistics, the lack of higher redshift quasars, and the coarseness of the Gunn-Peterson test make that inference

**Table 1:** *Predicted number of high-redshift quasars for double power-law luminosity function with $\beta_1$=-1.64, $\beta_2$=-3.2, M\*(1450Å)=-24.5, and $6 \times 10^{-10}$ Mpc$^{-3}$ quasars brighter than -26.7 at z=6 (Fan et al. 2004; see Prochaska et al. 2009). We assume redshift evolution proportional to exp(-0.43z), normalized to SDSS results at z=6, and require M(1450)<-23.*

| Redshift | Lyα Break (μm) | NIRSS | UKIDSS |
|---|---|---|---|
| 6 < z < 8 | 0.9 – 1.1 | ~100,000 | 70 – 150 |
| 8 < z < 10 | 1.1 – 1.3 | ~35,000 | 1.3 – 2.6 |
| z > 12 | > 1.6 | ~7,000 | 0 |
| z > 20 | > 2.6 | ~100 | 0 |

somewhat uncertain (*e.g.*, Stern *et al.* 2005).

Currently, the most distant known quasar is at *z* = 6.4 (Fan *et al.* 2001), and only a dozen quasars have been confirmed at *z* > 6 (*e.g.*, Stern *et al.* 2007). Ambitious new surveys seek to push this number to around 100, identifying one or two quasars at *z* ~ 8 (Table 1; *c.f.*, Fontanot *et al.* 2007). *NIRSS* will <u>fundamentally</u> change the landscape of early universe investigations. Based on the Fan *et al.* (2001, 2004) quasar luminosity functions, *NIRSS* will identify tens of thousands of quasars at *z* > 6, hundreds of quasars at *z* > 15, and push out to *z* > 20 should quasars exist at those redshifts. Such discoveries will directly measure the first epoch of supermassive black hole formation in the universe, probe the earliest phases of structure formation, and provide unique probes of the intergalactic medium along our line of sight to these distant, luminous sources. In particular, the large numbers of quasars identified in the first Gyr after the Big Bang will enable clustering analyses of their spatial distribution, which directly probes the mean mass of their dark matter halos (*e.g.*, Coil *et al.* 2007, Myers *et al.* 2007).

The strong spectral signature from redshifted Lyman-alpha forest absorption will make high-redshift quasars easily identifiable from the *NIRSS* data alone. Moreover, optical surveys (e.g., SDSS, LSST, Pan-STARRS) are not capable of identifying quasars above *z* ~ 6.5 since their optical light is completely suppressed by the redshifted Lyman break and Lyman forests. Similarly, JDEM will not be sensitive to quasars at *z*>12. *NIRSS* will be the definitive probe of the first phase of supermassive black hole growth in the universe.

## 2.3 The Stellar and Dark Matter Content of Galaxies

A primary goal of the next generation of deep, ground-based optical surveys is to study the nature of dark matter via weak lensing. Such analysis uncovers the underlying three-dimensional distribution of dark matter in the Universe and studies its evolving relationship to luminous matter. Weak lensing studies require both (1) excellent image quality in order to study correlated distortions in image shape parameters due to foreground mass, and (2) excellent photometric redshifts in order to separate sources into redshift bins and differentiate foreground and background sources. While the expected performance of *NIRSS* should be competitive for the former, the main advantage of *NIRSS* will be the longer wavelength data which will significantly enhance photometric redshifts at *z*>1. As shown in Fig. 3, *NIRSS* photometry combined with deep optical data from LSST will improve photometric redshifts for well-detected galaxies at *z*>1 by nearly a factor of ten, thereby enabling tomographic mapping of the large scale distribution of dark matter in the universe to earlier cosmic epochs.

Furthermore, while the rest-frame UV light of galaxies comes primarily from young stars and is a good measure of the current star formation rate, the rest-frame optical and near-infrared comes primarily from lower mass stars and is thus a good measure of the integrated star formation rate, *e.g.*, the current stellar mass of a galaxy. Thus *NIRSS* will probe both the stellar and dark matter content of the universe over the past 10 Gyr. *NIRSS* will test theories of structure formation out to $z\sim4$, significantly amplifying the scientific capabilities of the optical surveys alone.

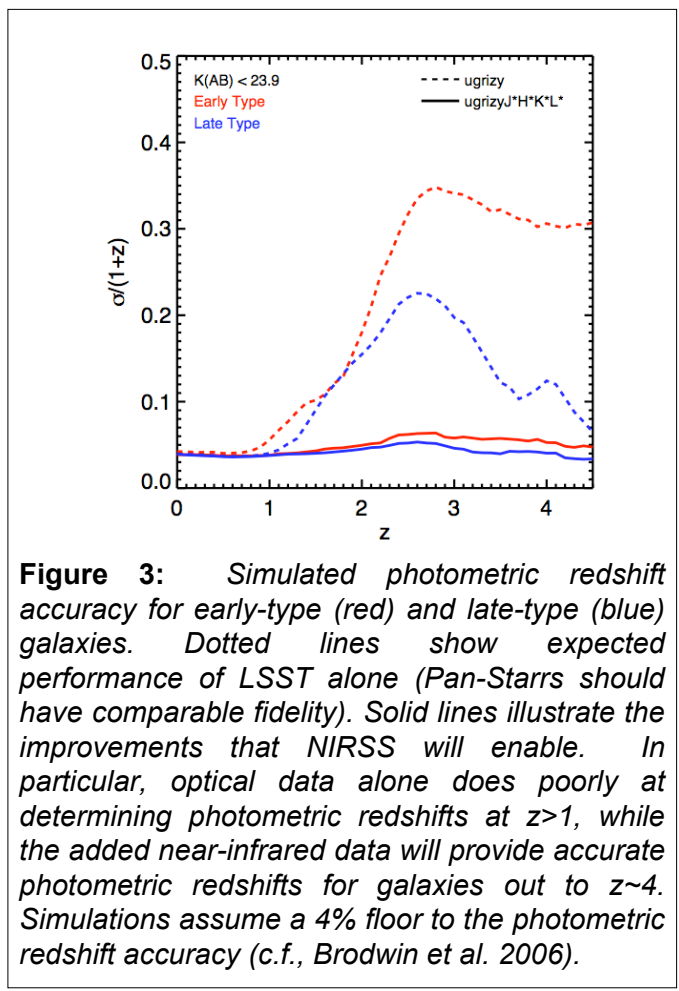

**Figure 3:** *Simulated photometric redshift accuracy for early-type (red) and late-type (blue) galaxies. Dotted lines show expected performance of LSST alone (Pan-Starrs should have comparable fidelity). Solid lines illustrate the improvements that NIRSS will enable. In particular, optical data alone does poorly at determining photometric redshifts at z>1, while the added near-infrared data will provide accurate photometric redshifts for galaxies out to z~4. Simulations assume a 4% floor to the photometric redshift accuracy (c.f., Brodwin et al. 2006).*

## 2.4  The Coldest Stars

### 2.4.1  Low Mass Hydrogen Burning Stars and Brown Dwarfs

In addition to the extragalactic science goals, *NIRSS* is ideally suited for several Galactic studies. With a depth of 25.6 mag (AB) in multiple near-infrared passbands, *NIRSS* will provide a complete census of low mass hydrogen burning stars in the Solar neighborhood. At an age of 1 Gyr, a 0.08 Solar mass star has $M_V = 19$, but a color of $V\text{-}H = 8.3$ (Baraffe *et al.* 1998). Therefore, although this population of stars cannot be seen at large distances in the optical, we will easily identify and study the properties of these stars out to 5 kpc in the near infrared.

The unprecedented full-sky sensitivity of *NIRSS* will also extend Galactic studies to sub-stellar objects, probing below the hydrogen-burning limit. Observationally, the most important topic in this field is the characterization of objects cooler than T dwarfs (temperatures less than 700 K) – so called "*Y dwarfs*". Such objects have spectra that peak in the infrared (*e.g.*, Burrows *et al.* 2003) making them very faint at optical wavelengths. *NIRSS* will identify large samples of the coldest brown dwarfs, thus enabling study of their distribution within the Galaxy. For example, while a 600 K brown dwarf could only be detected by 2MASS to ~1 pc, *NIRSS* will see such sources out to ~50 pc. Multi-epoch *NIRSS* data will also provide proper motions down to 0.05" per year (per axis), which will identify any dwarf with tangential velocity > 10 km/s out to 25 pc. Because our goal is to probe into the uncharted territory of Y dwarfs, it is difficult to predict the number of brown dwarfs *NIRSS* will identify. However, the models of Martin *et al.* (2001) predict approximately 2000 Y dwarfs would be identified by *NIRSS*. *NIRSS* will provide a crucial test for the structure and evolution of both low mass hydrogen burning stars and brown dwarfs. The results will lead to an improved estimate of the Galactic mass budget, the initial mass function, and the threshold mass signifying hydrogen versus deuterium burning.

Besides *NIRSS* studies of cool stars in the field, the complete coverage of *NIRSS* will also provide such data for many star clusters, probing far below the hydrogen-burning limit. Such data is especially valuable as cluster members are coeval and have similar metal content. An ultra-deep, space-based study of the second nearest globular cluster has recently revealed stars near the hydrogen-burning limit (Richer *et al.* 2006). Our survey will extend this detection to other globular clusters (*e.g.*, M4, 47 Tuc, and NGC 6752), as well as to a large sample of nearby open star clusters (*e.g.*, Kalirai *et al.* 2001). As these systems have a large range in both age and metallicity, this study will improve understanding of how the initial mass function and hydrogen-burning limit vary with environment (*e.g.*, Kroupa 2001).

### 2.4.2  Halo White Dwarfs

Over 97% of all hydrogen burning stars will eventually exhaust their fuel and cool to become white dwarf stars. Given the age of the Galactic halo, most of the mass in this component is now tied up in these remnant stars (*e.g.*, Alcock *et al.* 2000). White dwarfs are initially very hot and blue, but become cooler and fainter with time. The formation of molecular hydrogen in hydrogen atmosphere white dwarfs causes them to become bluer when their temperatures drop below 5000 K (Hansen 1999). However, helium atmosphere white dwarfs continue to become redder as they cool. At an age of 10 Gyr, a helium atmosphere white dwarf will have cooled to 3500 K, will be quite faint optically ($M_V = 18$), but will have a red optical-to-near-IR color, $V$-$H = 4$ (Bergeron, Saumon & Wesemael 1995). Therefore, *NIRSS* will see the entire population of these cool Helium white dwarfs out to 1.1 kpc. At optical wavelengths, only the deepest pencil beam surveys are able to probe this population of stars out to this distance. *NIRSS* will derive the helium white dwarf luminosity function, which will directly yield independent ages for the Galactic halo and disk (*e.g.*, Winget *et al.* 1987; Ferrario *et al.* 2005).

## 2.5  Microlensing Survey for Exoplanets

The Exoplanet Task Force (Lunine *et al.* 2008) expressed strong support for space-based microlensing studies of exoplanets, which can be achieved with a 1-meter aperture space telescope with a greater than 0.5 deg$^2$ field of view and an orbit allowing continual access to the Galactic bulge. Microlensing is most sensitive at planetary separations of ~$R_E$ from the lens star, provides distinct and diagnostic light curves, and produces strong, easily detected (~10%) photometric variations (Bennett *et al.* 2009). Microlensing is sensitive to very low mass planets (~0.1 $M_E$) and large semi-major axes (~1 – 10 AU), and is thus complimentary to the exoplanet discovery space probed by Doppler studies, *Kepler* and *SIM*. The *NIRSS* payload is quite similar to that of the *Microlensing Planet Finder*, a 1.1 meter near-infrared mission concept with a 1.3 deg$^2$ focal plane populated with 35 Hawaii-2RG detectors (Bennett *et al.* 2004). Scaling from that proposed mission and estimates from the Exoplanet Task Force, allocating ~10% of the *NIRSS* 4-yr mission lifetime to synoptic observations of the Galactic bulge will have minimal impact on the proposed sky survey depths but would detect several Earth-mass planets in the habitable zone, as well as dozens to hundreds of gas giants at large (>1 AU) orbital radii. These latter systems are not found by eclipsing missions such as *Kepler*.

## 2.6  Why this depth?  Why now?  Why not from the ground?

The full-sky, 0.2 µJy-depth, near-infrared images provided by *NIRSS* will produce significant science that is unattainable with current or planned projects. This depth is well matched to the next generation of planned optical wide-area (> $2\pi$ ster) surveys and will significantly enhance the photometric redshift fidelity for sources identified in those surveys. The long wavelength data of *NIRSS* will also complement the optical data to provide wide-band spectral energy distributions of billions of sources. Shallower near-IR data would not achieve sufficient photometric redshift fidelity for sub-L* galaxies at $z$>1, while significantly deeper data is not necessary. Since these optical surveys — as well as the launch of *JWST* — are expected to occur midway through the next decade, launching *NIRSS* in the same period will provide maximal synergy with those projects. In particular, *NIRSS* will provide an ideal survey to identify targets for both *JWST* and the next generation of extremely large, 30-m class ground-based telescopes. Finally, the area and depth planned by *NIRSS* are not feasible from the ground where the thermal environment produces an infrared background ~1000 times brighter, in addition to having spatially and temporally variable water absorption and OH and thermal emission. Scaling from the VISTA and UKIDSS programs, which will be shallower than *NIRSS* over <0.1% of the sky using dedicated multi-year surveys on 4-meter class telescopes, full-sky near-infrared images at the 0.2 µJy depth would require a network of hundreds of 4-m class telescopes operating for nearly a decade. With the expense of the telescopes, each having its own expensive near-infrared focal plane, such an option is not cost competitive with a space-based mission. Furthermore, the longest wavelength channel of *NIRSS* would not be feasible from the ground. The scientific gains from *NIRSS* will be diverse, providing a fundamental legacy database useful to the entire astrophysical community (professional and amateur) for decades to come. As has been demonstrated with both the 2MASS and SDSS data sets, we expect heavy and unpredicted uses of such a sky survey reaching such unprecedented depths.

## 3  Technology Overview

The *NIRSS* mission builds on a well-established design and technology base. All elements of the architecture are well within what has been validated for missions such as *IRAS*, *Spitzer*, *WISE*, and *Kepler*. The concept does not require any low technology readiness level (TRL) technologies. In addition, many design elements have already been extensively studied for Joint Dark Energy Mission (JDEM) concepts such as *SNAP*, *DESTINY*, and *ADEPT*. While the science goals of *NIRSS* are quite different and the coverage extends to longer wavelengths, we will reduce cost and risk by leveraging the substantial research and development efforts already committed to these missions. No new spacecraft technology will be required.

The major new element is the large infrared focal plane. Infrared hybrid focal plane arrays (FPAs) have made major advances in recent years — exemplified by the Teledyne Hawaii-2RG array and the FPA for *JWST* NIRCam, which incorporates four Hawaii-2RG devices (§3.2). We will build on these developments to define the large focal plane required for *NIRSS*.

Compared to other astronomy missions operating at longer wavelengths, the FPA cooling requirements are modest. For an L2 orbit, passive cooling will get us to our required focal plane temperature of 45 K. *Spitzer*, which is in an Earth-trailing orbit, passively cools to below 40 K. Should thermal issues become significant, the additional cooling required is well within the capabilities of modern mechanical coolers.

While all-sky survey predecessors *IRAS* and *WISE* both used cryostats in low Earth orbit (LEO), the baseline *NIRSS* mission definition, described next, places it in an L2 orbit. Because smaller launch vehicles (such as the Delta-II) are not expected to be available, the cost penalty of going beyond LEO is significantly reduced.

### 3.1  Baseline Mission Definition

*NIRSS* will utilize a 1.5 meter f/10.6 telescope operating at L2, and will image the sky in four bandpasses ranging from 1.0 to 4.1 microns. For ease of reference to traditional ground-based filters, we call these bandpasses $J^*$ (1.2 μm, Δλ=0.4 μm), $H^*$ (1.6 μm, Δλ=0.4 μm), $K^*$ (2.2 μm, Δλ=0.8 μm) and $L^*$ (3.4 μm, Δλ=1.6 μm). Note that the actual filter shapes, free from the restrictions imposed by the Earth's atmosphere, are wider. The orbit, similar to that planned for *JWST* and JDEM, provides a thermally cool environment, thereby negating the need for a cryostat or mechanical cooler.

Baseline parameters are listed in Table 2. The *NIRSS* focal plane (Fig. 4) will be populated with 36 2048x2048 arrays, with a pixel scale of 0.25". These arrays will be packaged in nine modules (each effectively 4096x4096) similar to those being built for *JWST*. The pixel scale will slightly undersample the point spread function (PSF) in the shorter passbands — *e.g.*, the Rayleigh criterion diffraction limit at 1.2 μm is 0.20". Filters will be fixed to the arrays. The survey strategy will shift the telescope by 8.5 arcmin between 60 sec exposures along an array axis (*e.g.*, the extent of a single 2048x2048 array). After a sequence of exposures, including shifts in the orthogonal array axis, this survey strategy will image each point in the sky 24 times, evenly split between the four filters and two epochs. The three images thus obtained per epoch per filter provides robustness against bad pixels and cosmic rays. The temporal lags within this strategy will enable asteroid identification. After four years of operations, we will reach the proposed depth of ~0.2 µJy (5σ; point source) in all four near-infrared bands across the entire sky. As is clear from Fig. 1, confusion will not be an issue for *NIRSS*, at least for extragalactic fields. Low Galactic latitude fields will likely require a modified (shallower) observing strategy.

**Table 2:** *Baseline NIRSS parameters.*

| | |
|---|---|
| Primary Mirror Diameter: | 1.5 m (produces 0.20" resolution at 1.2 μm) |
| Pixel Scale: | 0.25 arcsec/pixel |
| Wavelength Range: | 4 bands, 1.0 – 4.2 μm |
| Survey Depth: | 0.2 μJy, full-sky |
| Optical Design: | f/10.6 TMA |
| Payload Mass: | 1400 kg |
| Focal Plane Power: | 22 W |
| Launch Vehicle / Orbit: | Delta IV / L2 orbit |
| Mission Duration: | 4 years |

**Table 3:** *Baseline FPA parameters.*

| Pixel Size: | 18 µm |
|---|---|
| FPA Temperature: | 45 K |
| Readnoise: | 8 e$^-$ (λ < 2.5 µm) <br> 15 e$^-$ (λ > 2.5 µm) |
| Dark Current: | 0.003 e$^-$/s (λ < 2.5 µm) <br> 0.01 e$^-$/s (λ > 2.5 µm) |

## 3.2  Focal Plane Arrays (FPAs)

The required spectral sensitivity is the primary consideration for selecting the detector material. At the wavelengths planned for *NIRSS*, HgCdTe or InGaAs are the most mature technologies. HgCdTe, the likely design choice for *NIRSS*, is included in our baseline design.

The *NIRSS* design is based on focal plane arrays identical to those under development for *JWST*. The primary technology driver for the *NIRSS* mission will be optimizing the architecture for the large focal plane based on the existing or achievable technology. There are at least two major considerations in the design of a large mosaicked focal plane array: (1) design of the individual devices for close packing (butting) and (2) ability to affordably produce the devices with the required performance. We assume the availability of a module on the order of 4000 elements square.

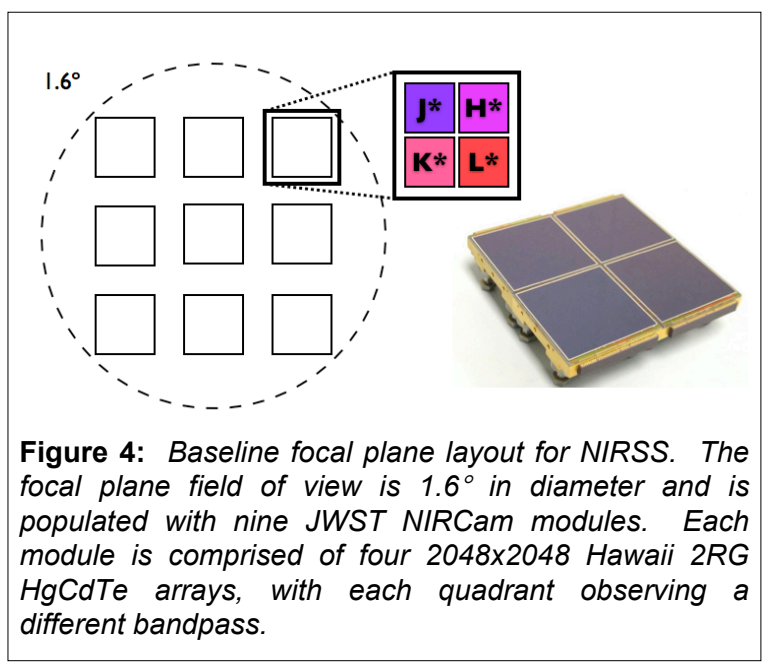

**Figure 4:** *Baseline focal plane layout for NIRSS. The focal plane field of view is 1.6° in diameter and is populated with nine JWST NIRCam modules. Each module is comprised of four 2048x2048 Hawaii 2RG HgCdTe arrays, with each quadrant observing a different bandpass.*

Fig. 4 provides one concept for how the *NIRSS* focal plane could be implemented. We have assumed the *JWST* NIRCam modular design. Table 3 summarizes the parameters of the Teledyne Hawaii-2RG array operating at 45 K, based on the performance capabilities of the NIRCam detectors using Fowler-8 sampling (Garnett *et al*. 2004, Beletic *et al*. 2008).

## 3.3  Thermal Design

Of the available cryogenic systems, passive cooling with multi-stage radiative surfaces is the most attractive method to cool *NIRSS*. Passive coolers are simple, reliable, low mass, require no power and have long lifetimes. They have been used extensively in applications requiring cooling in the 100-200 K range in LEO and can reach lower temperatures with the advantageous thermal field of view (FOV) to cold space afforded by an L2 or Earth-trailing orbit.

The ability to reject heat to space is limited by the radiator emissive power at its operating temperature. The emissive power at 70 K is 1.36 W/m$^2$. The net cooling power available is the balance of the emissive power minus the parasitic radiative and conductive heat loads from the warm support structure. In most applications below 100 K, the parasitic heat loads dominate. This typically favors a more efficient two-stage passive cooler design in which the first stage provides cooling to a thermal shield that surrounds the second stage. The thermal shield serves to protect the second stage from the spacecraft, the Sun and the Earth. The second stage provides cooling to the cold cryogenic components. There is substantial heritage in three or four passive cooler stages to achieve the desired temperature levels. For example, the outer shell of *Spitzer*, which is in an Earth-trailing orbit, is passively cooled to below 40 K.

JPL has extensive experience with multi-stage passive cooler designs such as the *Spitzer*, PMIRR, VIMS, *AIRS*, TES and M3 instruments. Both *AIRS* and TES have large two-stage passive coolers with radiating areas near ~1 m$^2$ and have been operating in space for over four years at an altitude of 705 km in a Sun-synchronous Earth orbit. The M3 instrument, launched in October 2008 onboard the *Chandrayaan-1* spacecraft and currently orbiting the Moon at an altitude of 100 km, makes use of a three-stage passive cooler design to cool its focal plane and optics. An important constraint for cryogenic passive cooling is that the spacecraft must have a single solar array panel on the Sun-side of the spacecraft and the passive radiator needs a clear FOV to cold space (*e.g.,* Fig. 5). Though the baseline design of *NIRSS* uses multi-stage passive cooling, we note that cryocoolers have a high level of technical readiness were additional cooling to become necessary (*e.g.*, passive would likely be insufficient were we to move to LEO).

### 3.4  Optical Design

The performance required of the *NIRSS* collection optics falls well within the capabilities of a class of telescopes known as three mirror anastigmats (TMAs). TMAs are superior to two-mirror telescopes for applications requiring both a large aperture and a large field of view. In fact, as the desired field size of a large aperture telescope approaches one degree, TMAs become the only credible design choice.

As three-mirror optical systems, TMAs are structurally more complicated, heavier and more expensive than two-mirror all-reflecting telescopes of comparable size. However, workable TMA structural designs are readily available and the additional mass and cost of TMAs are typically measured in percentages, not multiplicative factors. TMAs are therefore the design of choice for large aperture, wide field, space-based telescopes, *e.g.* those of the JDEM, *Ikonos-2* and *GeoEye-1* missions. The commercial *GeoEye-1*, which launched in September 2008, employs a TMA with a primary diameter of 1.1 m, effective focal length of 13.3 m, angular resolution better than 0.4 arcsec and field of view greater than 1.28 degrees. The *NIRSS* TMA is representative of current practice and poses only those risks generally associated with high-performance space-based optics.

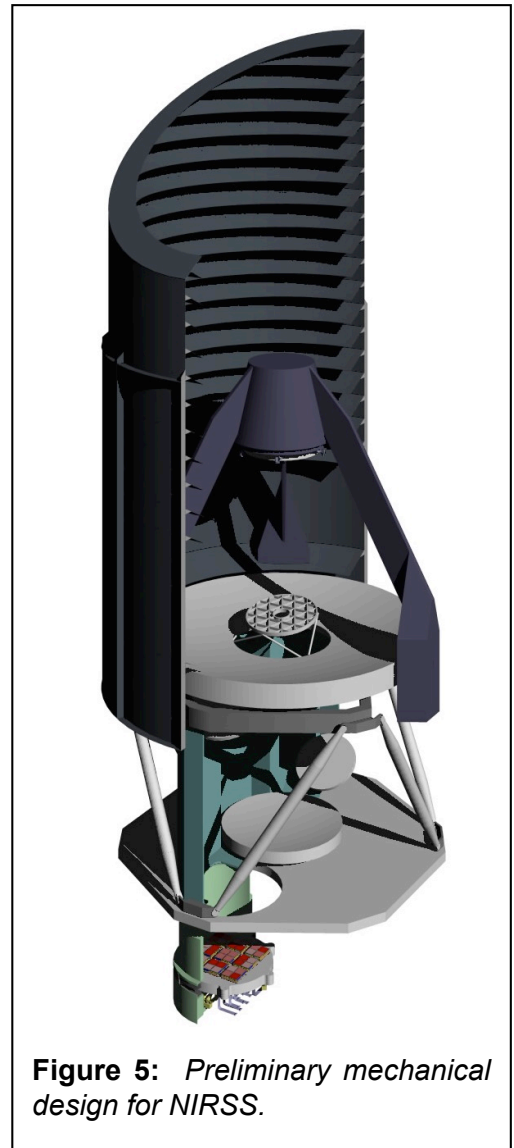

**Figure 5:** *Preliminary mechanical design for NIRSS.*

Fig. 6 presents the *NIRSS* optical design, which consists of a TMA with three additional fold mirrors. Redward of 1 μm, this design achieves diffraction-limited performance across the entire 1.6° FOV. This design is telecentric, *e.g.,* the chief rays are near normal to the image plane, which is advantageous for the interference filters. Relative to, for example, the *SNAP* design, the *NIRSS* design uses two additional fold mirrors in order to avoid a blind spot in the center of the FOV. The loss in throughput from these additional reflections is minimal, *e.g.,* protected gold has ~98% reflectance for the wavelength range 0.65 – 16 μm.

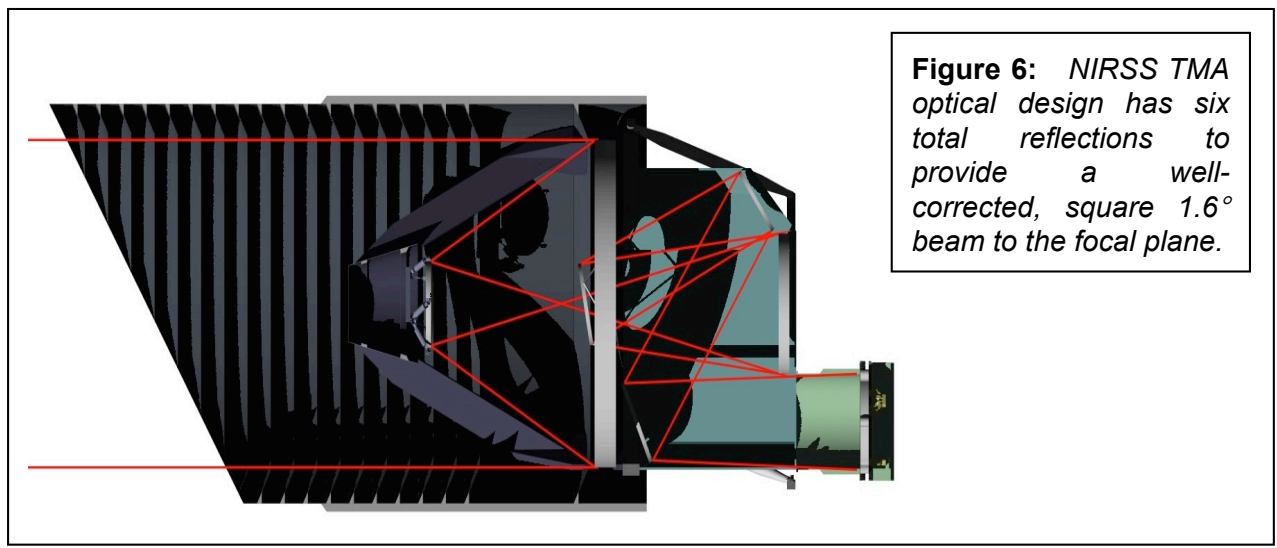

**Figure 6:** *NIRSS TMA optical design has six total reflections to provide a well-corrected, square 1.6° beam to the focal plane.*

### 3.5  Spacecraft and Mission Design

Thermal and pointing requirements will be the primary drivers of the spacecraft design. The requirement for the *NIRSS* payload to be kept at 70 K during its entire four-year nominal operational lifetime is the major motivation for an L2 orbit. An L2 orbit provides a stable thermal environment in which the Earth and Sun can both be kept out of the FOV of the radiator with relative ease, eliminating the need for active cooling. L2 orbits similarly simplify the task of minimizing stray light in the telescope, and can be designed such that eclipses occur rarely, or in the case of high-Z amplitude halo orbits, never occur. One of the major drawbacks to L2 orbits is the larger and more energy consumptive telecommunications hardware needed to downlink science data. However, upgrades to the DSN for the *JWST* mission are expected to significantly mitigate this issue by using 150 Mbps Ka-band downlinks with a 34-meter dish.

Pointing requirements on the *NIRSS* spacecraft are expected to be tight, but well within the range of what has been previously demonstrated on-orbit. A pointing stability of 0.1 arcsec will allow for the 0.2 μJy science goal. This pointing stability is comparable to the *Spitzer*'s pointing stability of <0.1 arcsec (1σ radial rms over 200 sec) and should prove readily achievable during the proposed 60 sec integrations. The *Kepler* and JDEM pointing stability requirements are typically five to ten times tighter.

The proposed step-stare survey method will allow for optimization of the pointing control system to accomplish 8.5 arcmin slews (*e.g.,* one 2Kx2K array width) in 10 sec. Using *Spitzer* as a conservative baseline for step-stare operations, the required slew rate of 50 arcsec/sec should be quite achievable. Settle times are expected to be on the order of 20 sec; *e.g.*, *Spitzer* settles to within 0.2 arcsec rms within 10 sec for slews less than 30 arcmin.

Details of the mapping strategy remain to be worked out and will depend upon the achievable detector abuttability, FPA layout design, and thermal viewing constraints. In particular, the requirement to maintain the shield pointed towards the Sun and the radiators pointed at cold space will tightly constrain observation angles. With an L2 orbit, *NIRSS* will drift by ~1 degree/day around the Sun. By slowly rotating the craft about the Sun-spacecraft line, *NIRSS* will use the heliocentric orbit to fill in the sky over time.

We also propose to dedicate a portion of the *NIRSS* mission lifetime to directed observations, likely allocated in a GO mode, thus enabling significant new science such as microlensing exoplanet searches towards the Galactic bulge. As another example, while *NIRSS* will easily see red giants in Andromeda and red supergiants in the Virgo cluster, the 0.2 µJy sensitivity is slightly too shallow to detect globular clusters in the Coma cluster. Such detections could easily be achieved with a small (~1 day) survey.

For the baseline design, *NIRSS* has 9 Hawaii-2RG detector arrays observing thru each filter at any time, covering a total area of (9 arrays) x (2048 x 2048 pix/array) x (0.25" /pix)$^2$ = 0.182 deg$^2$ per filter. With 90 sec per observation (integration + slew + settle time) and a requirement to observe each sky position at least three times per epoch in order to control for systematic noise contributions such as cosmic rays and bad pixels, we will require 1.95 yr of continuous observations to observe the full 41,253 deg$^2$ of the full sky. This calculation is aggressive in that it ignores efficiency losses due to data downlinks and safing events, though it also ignores the shorter exposure times that will be required at low Galactic latitudes in order to avoid confusion. Thus, *NIRSS* will require ~2 yr per epoch to map the full sky.

With two bytes per pixel, a single, full *NIRSS* FPA image is ~0.3 GB. Assuming observations halt during downlink, the slew plus settle time of 30 sec and integration time of 60 sec allows for ~6000 images per week, or nearly 2 TB of data per week. With 50% data compression and a 125-150 Mbps Ka-band data rate, science data can be downlinked in two ~8 hr tracks per week.

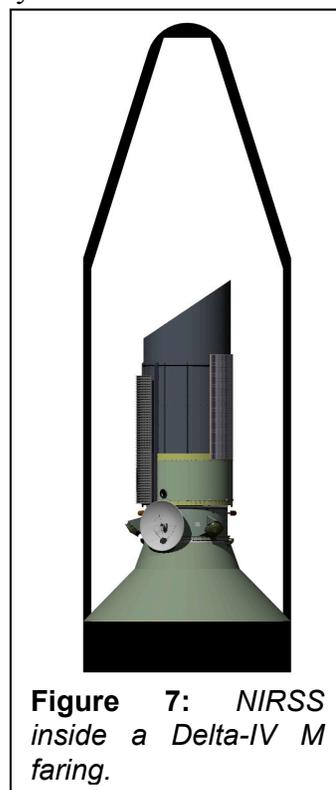

**Figure 7:** *NIRSS inside a Delta-IV M faring.*

### 3.6 Mass, Power, and Launch Vehicle

We estimate a total mass of 1400 kg for *NIRSS*, roughly divided between the telescope (650 kg), the instrument (230 kg) and the spacecraft (520 kg). Several launch vehicles have the capability to launch a satellite of this mass to an L2 orbit with significant margin. For example, Table 2 baselines the Delta-IV M which has a 10.3 m x 3.7 m diameter faring and has a maximum payload of 2780 kg for an L2 orbit. As seen in Fig.7, these parameters are ample for *NIRSS*. The Atlas V 401 and Falcon 9 are two other viable launch vehicles for *NIRSS*.

In terms of power, each Hawaii-2RG array has a detector-only power of 0.15 mW per output, and can be read with 1, 2, 4, or 32 outputs. We assume the 4-output read mode which produces an acceptable 0.6 mW per detector thermal load on the focal plane, or 19.2 mW in total. An additional ~700 mW per array is required by the SIDECAR electronics (which is thermally isolated from the focal plane), implying a total power consumption of ~22 W for the focal plane.

# 4 Technology Drivers

## 4.1 Large Near-Infrared Focal Plane

The *NIRSS* mission concept entails no major technological hurdles. All aspects of the *NIRSS* mission are well within the range of what has been validated for previous missions such as *IRAS*, *Spitzer*, *Kepler*, and *WISE*. With funding, a relatively rapid design phase could proceed rapidly into construction, launch, and operations.

**Table 4:** *Noise per resolution element per 60s exposure.*

|  | J* | H* | K* | L* |
|---|---|---|---|---|
| pix/resolution element: | 1 | 2 | 3 | 6 |
| zodi intensity (MJy/sr): | 0.15 | 0.13 | 0.10 | 0.08 |
| σ(readnoise): | 8.0 | 11.3 | 13.9 | 36.7 |
| σ(dark current): | 0.2 | 0.3 | 0.3 | 1.5 |
| σ(zodi): | 7.6 | 8.7 | 11.1 | 16.4 |

*Average zodiacal dust intensities are from a DIRBE-based model assuming 90° solar elongation and 50° ecliptic latitude. Zodi intensity will vary by a factor of ~3-5 across the sky and represents both reflected sunlight (at shorter wavelength) and thermal zodi emission (at longer wavelength). Numbers assume the baseline mission, gain=1 and 50% throughput.*

The one slightly challenging technology driver for *NIRSS* is the large near-infrared focal plane. As envisioned, the *NIRSS* focal plane will be populated with 36 Hawaii-2RG detectors with capabilities matching the *JWST* NIRCam detectors (Garnett *et al.* 2004). This focal plane, though heavily-populated, is similar to that planned for the *Microlensing Planet Finder* and is significantly less aggressive than the *Kepler* focal plane. Comparable focal planes have also been considered for various JDEM concepts such as *SNAP* and *Euclid*. Furthermore, focal plane size scales inversely with mission lifetime, providing a open trade study.

For much of the high ecliptic latitude sky, even the impressive NIRCam detectors will provide *NIRSS* with readnoise-limited observations in the short 60 s observations from which the sky survey is constructed (Table 4). A few paths towards the preferred situation of zodi-limited observations are:

1. *Development of HgCdTe detectors with lower readnoise.* A 20% improvement in readnoise capabilities would provide zodi-limited observations for much of the celestial sphere for most of the *NIRSS* passbands. Dark current is unlikely to be a driving noise term, so if detectors and/or readout modes could be developed with lower readnoise at the cost of higher dark current, that would aid in the goal of reaching zodi-limited data.
2. *Longer exposure times.* The 60 s individual exposure time currently planned is set by the requirement to observe the full sky in approximately two years with a focal plane populated by 36 2k x 2k detectors observing four bands with pixel scales matched to the diffraction limit of the telescope. Three exposures per sky position per bandpass per epoch are baselined in order to control systematic errors. Implementing *NIRSS* with more pixels (and keeping all other major mission parameters fixed) would allow longer dwell times per sky position. More pixels could come from a combination of detectors, more detectors, and/or improvement in the detector abuttability. However, the addition of detectors raises concerns regarding thermal loads on the satellite.
3. *Larger pixel scales.* This is not recommended since we are already undersampling the PSF at most *NIRSS* passbands.

# 5   Activity Organization, Partnerships, Status

## 5.1   Organization, Roles, and Responsibilities

**Principal Investigator**

Daniel Stern of JPL will manage *NIRSS*, including all reporting to NASA, and will directly manage the activities of all co-investigators. He will delegate day-to-day management of the engineering team to the Project Manager.

**Science Team Co-Investigators**

The science team has extensive experience in large infrared surveys from space and from the ground. The science team includes several senior scientists from the *WISE* mission, including the mission PI (Wright), the Project Scientist (Eisenhardt) and the Deputy Project Scientist (Mainzer). Data from both *WISE* and *NIRSS* will be processed by IPAC, the Deputy Executive Director of which is co-I Cutri. Cutri was also Project Scientist for 2MASS, the ground-based predecessor to *NIRSS*. Co-I's Dey, Eisenhardt, Gonzalez and Stern have each led synergistic deep wide-area, infrared surveys of the Boötes field. These surveys involved co-I's Brodwin, Cooray, Moustakas and Stanford. In particular, Brodwin is an expert in photometric redshifts derived from multi-wavelength data sets, while Cooray is a theoretical cosmologist with an expertise in diffuse backgrounds. Co-I Rhodes is an expert in weak gravitational lensing and is an active co-I in the *SNAP* and *Euclid* dark energy mission concepts. He will develop the synergism between *NIRSS* and the next generation of deep optical surveys. Co-I Kalirai is an expert in a wide range of Galactic and local group investigations, including studying cool stars, white dwarfs, open clusters, globular clusters, and structure in the Andromeda galaxy. He will develop the Galactic and local universe science case of *NIRSS* with co-I Mainzer.

Primary issues for the science team will be defining the optimal filter selection, optimal survey strategy, and determining quantitative metrics between various options. All team members will be involved in these fundamental issues, with co-I's contributing according to their area of expertise and field of interest.

## 5.2   Partnerships

No substantial partnerships beyond the science team have yet been established for the *NIRSS* mission concept. As currently envisioned, JPL would lead the project management. Detectors would likely be procured from Teledyne, though no official price quotes have been requested thus far. Data processing and public release would be through IPAC, led by co-I Cutri.

## 5.3   Status

*NIRSS* is a new mission concept, still in the early days of formulation (*e.g.,* pre-Phase A). Many trade studies remain to be assessed. As currently presented, *NIRSS* is a large medium-class mission and reaches sensitivities equal to or better than the deepest pencil beam (*e.g.,* <10 square arcmin) surveys yet obtained, though *NIRSS* reaches these depths across the entire sky. Scaled back versions of *NIRSS* would still be quite interesting. In particular, and as advocated by P. Hertz in a presentation to the SEUS and OS in July 2003, we advocate that NASA fund a BIGEX mission class to complement the already fruitful SMEX and MIDEX programs (*e.g.,* Elvis *et al.* 2009). At $500M (without launch), many exciting scientific programs become available, which are not possible at the MIDEX or SMEX cost caps. The current launch vehicle environment also suggests this approach.

# 6   Activity Schedule

The *NIRSS* mission is relatively straightforward, with no significant technology drivers or engineering hurdles. As detailed in Table 5, assuming funding began at the start of FY11, we estimate that only two years, in total, would be required for Phases A and B, such that detailed design and construction could begin in FY13. We estimate that 3.5 years would be required for detailed design, construction, integration, and testing, such that launch could occur as early as 2016, which would provide significant overlap with the *JWST* mission. The baseline mission would last four years and would map the entire sky six times, reaching a total depth of ~0.2 µJy (5σ, point sources). The only long-lead item for *NIRSS* is the large number of detectors for the focal plane. However, with 5.5 years between Phase A and D, this does not introduce a significant schedule risk.

**Table 5:** *Phased activity schedule.*

| Phase | Length (years) |
|---|---|
| Phase A | 1 |
| Phase B | 1 |
| Phase C/D | 3.5 |
| Phase E | 4 |
| **Total** | **9.5** |

# 7 References (updated on 2010 August 19)